\documentclass[twocolumn,superscriptaddress,showpacs,preprintnumbers,amsmath,amssymb,prl]{revtex4}
\usepackage{graphicx}
\usepackage{dcolumn}
\usepackage{bm}

\begin{document}
\title{Operating Spin Echo in the Quantum Regime for an Atomic-Ensemble Quantum Memory}
\author{Jun Rui}
\affiliation{Hefei National Laboratory for Physical Sciences at Microscale and Department of Modern Physics, University of Science and Technology of China, Hefei, Anhui 230026, China}
\affiliation{CAS Center for Excellence and Synergetic Innovation Center in Quantum Information and Quantum Physics, University of Science and Technology of China, Hefei, Anhui 230026, China}
\author{Yan Jiang}
\affiliation{Hefei National Laboratory for Physical Sciences at Microscale and Department of Modern Physics, University of Science and Technology of China, Hefei, Anhui 230026, China}
\affiliation{CAS Center for Excellence and Synergetic Innovation Center in Quantum Information and Quantum Physics, University of Science and Technology of China, Hefei, Anhui 230026, China}
\author{Sheng-Jun Yang}
\affiliation{Hefei National Laboratory for Physical Sciences at Microscale and Department of Modern Physics, University of Science and Technology of China, Hefei, Anhui 230026, China}
\affiliation{CAS Center for Excellence and Synergetic Innovation Center in Quantum Information and Quantum Physics, University of Science and Technology of China, Hefei, Anhui 230026, China}
\author{Bo Zhao}
\affiliation{Hefei National Laboratory for Physical Sciences at Microscale and Department of Modern Physics, University of Science and Technology of China, Hefei, Anhui 230026, China}
\affiliation{CAS Center for Excellence and Synergetic Innovation Center in Quantum Information and Quantum Physics, University of Science and Technology of China, Hefei, Anhui 230026, China}
\author{Xiao-Hui Bao}
\affiliation{Hefei National Laboratory for Physical Sciences at Microscale and Department of Modern Physics, University of Science and Technology of China, Hefei, Anhui 230026, China}
\affiliation{CAS Center for Excellence and Synergetic Innovation Center in Quantum Information and Quantum Physics, University of Science and Technology of China, Hefei, Anhui 230026, China}
\author{Jian-Wei Pan}
\affiliation{Hefei National Laboratory for Physical Sciences at Microscale and Department of Modern Physics, University of Science and Technology of China, Hefei, Anhui 230026, China}
\affiliation{CAS Center for Excellence and Synergetic Innovation Center in Quantum Information and Quantum Physics, University of Science and Technology of China, Hefei, Anhui 230026, China}

\begin{abstract}
Spin echo is a powerful technique to extend atomic or nuclear coherence time by overcoming the dephasing due to inhomogeneous broadening. However, applying this technique to an ensemble-based quantum memory at single-quanta level remains challenging. In our experimental study we find that noise due to imperfection of the rephasing pulses is highly directional. By properly arranging the beam directions and optimizing the pulse fidelities, we have successfully managed to operate the spin echo technique in the quantum regime and observed nonclassical photon-photon correlations. In comparison to the case without applying the rephasing pulses, quantum memory lifetime is extended by 5 folds. Our work for the first time demonstrates the feasibility of harnessing the spin echo technique to extend lifetime of ensemble-based quantum memories at single-quanta level.
\end{abstract}


\maketitle

Recent years have witnessed remarkable progresses in the development of quantum memories with photonic interface. Many quantum systems \cite{Simon2010}, such as single neutral atoms \cite{Specht2011a,Hofmann2012}, single trapped ions \cite{Olmschenk2009,Stute2012}, single quantum dots \cite{DeGreve2012,Gao2012}, solid-state ensembles \cite{Tittel2010} and atomic-gas ensembles \cite{Sangouard2011} have been employed to store single photons or create entanglement with a single photon. Among them, the atomic-ensemble approach \cite{Gorshkov2007} is particularly attractive since the light-matter coupling is largely improved by collective enhancement. Plenty of important experimental progresses have been made in recent years \cite{Sangouard2011,Hammerer2010,Tittel2010}.

In an atomic-ensemble quantum memory, inhomogeneous broadening due to ambient magnetic field, atomic random motion and interaction with host spins etc. severely limits the storage time \cite{Zhao2009, Afzelius2010}. One universal solution to overcome inhomogeneous broadening induced decoherence is to make use of the spin echo technique \cite{Hahn1950}, where a series of $\pi $ pulses are applied to reverse the phase evolution through population inversion. This technique has been widely used in storage of classical light pulses. For example, with the spin echo technique, the storage lifetime has been extended to second and minute regime in solid-state ensembles and atomic-gas ensembles, respectively \cite{Longdell2005, Dudin2013}. However, whether this technique is applicable to the storage of quantum light or photons has not been resolved yet experimentally. The main concern \cite{Johnsson2004, Heshami2011} is that since the $\pi$ pulses induce population inversion, tiny imperfection of them could result in background noises which are much stronger than the stored single-photon signals.

In this letter we experimentally study the spin echo process of single excitations in a cold-atomic-gas quantum memory by employing stimulated Raman transitions. We find that the noise due to imperfection of the $\pi$ pulses is highly directional. In our experiment, by carefully arranging the Raman-beam directions and optimizing the pulse fidelities, we have successfully reduced this noise to much lower than the single-photon signal. Quantum nature of the spin echo process is verified by observing nonclassical photon-photon correlations. In our demonstration, the distorted spin-wave state gets rephased by applying two $\pi$ pulses and the quantum memory lifetime is increased by 5 folds. Our findings and techniques developed is applicable to all other ensemble-based quantum memories \cite{Simon2010}.


\begin{figure*}[hbtp]
\includegraphics[width=0.6\textwidth]{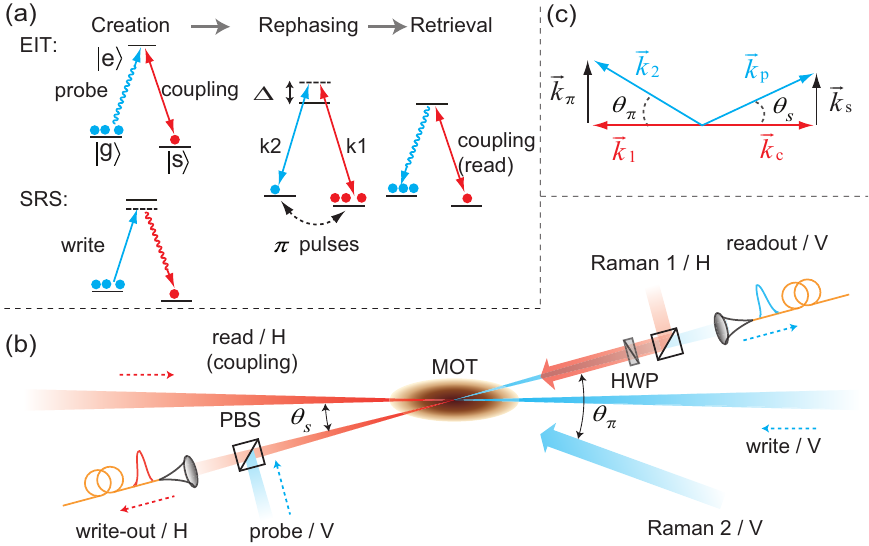}
\caption{(color online). Experimental setup. (a) Two methods are used in creating spin-waves in an atomic ensemble. In the electromagnetically induced transparency (EIT) process, a weak probe pulse at single-photon level is converted to atomic spin waves by turning off the coupling beam. In the spontaneous Raman scattering (SRS) process, a single-quanta spin-wave is imprinted in the atomic ensemble heralded by detecting a Raman scattered write-out photon. In the rephasing precess, rephasing pulses which couple the $|g\rangle\leftrightarrow|s\rangle$ transition through a two-photon Raman transition is applied. Later, spin-wave states are retrieved either by turning on the coupling or the read beam. (b) Schematic view of the experimental configuration. The atomic ensemble is prepared through magneto-optical trap (MOT). The coupling beam and the read beam have the same frequency, polarization and spatial mode, so do the the probe and read-out photon. $H$ ($V$) refers to horizontal (vertical) polarization relative to the drawing plane. HWP and PBS represents half-wave plate, and polarized beam-splitter, respectively. (c) Momentum relationships for the control beams, detection modes and the Raman beams.}
\label{fig1}
\end{figure*}

In an atomic-ensemble quantum memory, a single quantum state is stored as a spin wave spreading over the whole ensemble \cite{Fleischhauer2000,Duan2001}. The spin-wave state at $t=0$ can be described as
\begin{equation*}
|\Psi \rangle _{gs}=\frac{1}{\sqrt{N}}\sum_{j}^{N}e^{i\mathbf{k}_{s}\cdot
\mathbf{r}_{j}(0)}|g...s_{j}...g\rangle ,
\end{equation*}
where $N$ is the number of atoms, $|g\rangle $ and $|s\rangle $ are two atomic ground states, $\mathbf{k}_{s}$ is the wavevector of the spin wave, $\mathbf{r}_{j}(0)$ is the position of the $j$-th atom in the ensemble at $t=0 $. This state can be physically interpreted as a phase grating, which enables strong collective interference in the read-out process \cite{Duan2001}. With this collective interference, efficient conversion from spin waves to photons has been demonstrated \cite{Simon2007, Hedges2010, Bao2012}. As the spin wave is a collective exication over the whole ensemble, inhomogeneity of frequency or phase difference between the atomic levels of $|g\rangle$ and $|s\rangle$ for all atoms will distort the relative phase between each term in $|\Psi\rangle_{gs}$, thus results in a distorted phase grating. Atomic motion is one dominant decoherence mechanism for atomic-ensemble quantum memories \cite{Zhao2009, ZhaoR2009}. Within an atomic ensemble, the velocity $\mathbf{v}_j$ of each atom varies from atom to atom, which will distort the original phase grating in $|\Psi\rangle_{gs}$. After a storage time of $t=T$, the mismatching phase of $j$-th term in $|\Psi \rangle _{gs}$ is $\Delta \phi _{j}=\mathbf{k}_{s}\cdot \mathbf{r}_{j}(T)-\mathbf{k}_{s}\cdot \mathbf{r}_{j}(0)=\mathbf{k}_{s}\cdot \mathbf{v}_{j}T$ where we assume the atoms are freely moving with $\mathbf{r}_{j}(T)=\mathbf{r} _{j}(0)+ $ $\mathbf{v}_{j}T$. This results in a storage time \cite{Zhao2009} of $\tau_{s}\simeq1/k_{s}\bar{v}$ with $\bar{v}$ the average atomic thermal velocity.

This phase distortion can be eliminated by applying a spin-echo rephasing technique. As shown in Fig. \ref{fig1}, we apply two laser beams to induce stimulated Raman transitions \cite{Kasevich1991} between $|g\rangle $ and $|s\rangle $, where one laser couples the $|s\rangle \leftrightarrow |e\rangle $ transition with a wavevector of $\mathbf{k}_{1}$, and the other laser couples the $|g\rangle \leftrightarrow |e\rangle $ transition with a wavevector of $\mathbf{k}_{2}$. The rephasing scheme is implemented by applying two Raman $\pi $ pulses at $t=t_{1}$ and $t_{2}$ respectively. During the first $\pi $ pulse, an atom in the state $|g\rangle $ is transferred to $|s\rangle $ by absorbing a photon with momentum $\hbar \mathbf{k}_{2}$ and coherently emits a photon with momentum $\hbar \mathbf{k}_{1}$, thus obtains a phase of $\mathbf{k}_{\pi }\cdot \mathbf{r}(t_{1})$ with $\mathbf{k}_{\pi}=\mathbf{k}_{2}-\mathbf{k}_{1}$. While an atom in the state $|s\rangle $ is transferred to $|g\rangle $ by absorbing a photon with momentum $\hbar \mathbf{k}_{1}$ and coherently emits a photon with momentum $\hbar \mathbf{k}_{2}$, thus obtains a phase of $-\mathbf{k}_{\pi }\cdot \mathbf{r}(t_{1})$. Therefore, after the first $\pi $ pulse, the $j$-th term in $|\Psi \rangle _{gs}$ is changed from $|g...s_{j}...g\rangle $ to $|s...g_{j}...s\rangle $ and acquires a net phase $-2\mathbf{k}_{\pi }\cdot \mathbf{r}_{j}(t_{1})$, where we have neglected the overall phase. After the second $\pi $ pulse, the $j$-th term is transferred back to $|g...s_{j}...g\rangle $ and another phase of $2\mathbf{k}_{\pi }\cdot \mathbf{r}_{j}(t_{2})$ is obtained. Consequently, the overall phase gained by these two $\pi $ pulses is $\Delta \phi _{j}^{\pi}=2\mathbf{k}_{\pi }\cdot (\mathbf{r}_{j}(t_{2})-\mathbf{r}_{j}(t_{1}))=2\mathbf{k}_{\pi }\cdot \mathbf{v}_{j}\Delta t$ with the interval $\Delta t=t_{2}-t_{1}$. If $\Delta \phi _{j}^{\pi }$ is equal to $\Delta \phi _{j}$, the random phases cancel with each other and thus the spin-wave can be efficiently read out at $t=T$. In this way, we obtain the rephasing condition of $2\mathbf{k}_{\pi }\Delta t=\mathbf{k}_{s}T$, which sets critical constraints on the direction of Raman beams and the time interval between the two $\pi $ pulses, and the read out time $T$.

\begin{figure}[hbtp]
\includegraphics[width=\columnwidth]{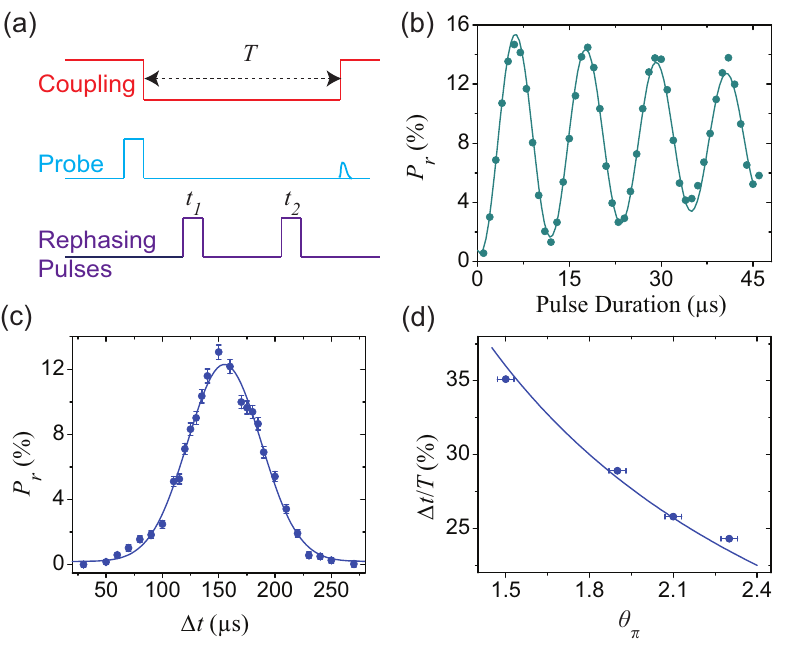}
\caption{(color online). Verification of the rephasing condition via EIT storage. (a) Time sequences for the coupling, probe and rephasing pulses. (b) Two-photon Raman Rabi oscillations between two ground states $|g\rangle$ and $|s\rangle$. The vertical axis is the photon detection probability in the read-out mode, which is proportional to the atom population in $|s\rangle$ state. (c) Optimization of time interval $\Delta t=t_2-t_1$ between two $\protect\pi$ pulses for a storage time of $T=600$ $\protect\mu$s. (d) Measured relationship between $\Delta t/T$ and the intersection angle $\protect\theta_\protect\protect\pi$ between the two Raman beams. The solid line refers to the theoretical estimate of $\theta_s/{2\theta_\pi}$ determined by the rephasing condition.}
\label{fig2}
\end{figure}

The layout of our experiment is shown in Fig.~\ref{fig1}. An ensemble of $\sim $10$^{8}$ $^{87}$Rb atoms are loaded in a magneto-optical trap. After polarization-gradient cooling, the temperature obtained is $\sim$10 $\mu $K, and the optical depth is $\sim $4. The energy levels employed are $|g\rangle \rightarrow |F=2,m_{F}=0\rangle $, $|s\rangle \rightarrow |F=1,m_{F}=0\rangle $ and $|e\rangle \rightarrow |F^{\prime }=2,m_{F}=\pm1\rangle $ of the D$1$ line. Note that by employing the ``clock states" $|F=2,m_{F}=0\rangle $ and $|F=1,m_{F}=0\rangle $, the decoherence due to magnetic field are suppressed and the inhomogeneous broadening due to atomic random motion is isolated for experimental study. Initially we prepare all atoms into the state of $|g\rangle $ through optical pumping, which increases the atom temperature to $\sim $15 $\mu $K. Two Raman beams with the same power of 2.5 mW originate from two separate diode lasers, which are phase locked to a frequency synthesizer at 6.8 GHz, i.e., the frequency separation between $|g\rangle $ and $|s\rangle $. Single-photon detuning $\Delta $ for both Raman beams is +750 MHz relative to $|e\rangle$. The wave number $k_{\pi } $ of Raman light can approximated by $k_{\pi }\approx k_{1}\theta _{\pi }$ if $\theta _{\pi }\ll 1$ (see Fig. 1c). In order to have high-fidelity Raman pulses, the Rabi frequencies for both Raman beams have to be stable for long time and identical for all the atoms. Therefore, we actively stabilize the intensity for both Raman beams with two independent digital proportional-integral controllers. We also increase the diameter of the Raman beams to 3.8 mm to improve the intensity homogeneity of the central region. Rabi flopping between $|g\rangle $ and $|s\rangle $ is measured as shown in Fig.~\ref{fig2}b. By fitting the curve with $I(\tau)=A \cos (2\pi \Omega _{r}\tau )e^{-\gamma \tau} + B$, we obtain a two-photon Rabi frequency of $\Omega _{r}=87.1$ kHz and a decay rate of $\gamma=13.4$ kHz. The fidelity of a single $\pi $ pulse is estimated to be $96\%$. Slight imperfection is mainly limited by slight intensity variations in the central region of the Raman beams, which originate from the imperfection of the optics used.

We first verify the rephasing condition via electromagnetically induced transparency (EIT) \cite{Fleischhauer2005}. A weak coherent laser pulse with an average photon number of $\sim $1 couples the $|g\rangle \leftrightarrow |e\rangle$ transition, and a control beam resonant with the transition of $|s\rangle\leftrightarrow |e\rangle $ controls the storage and read-out process. The waist of the probe beam is 90 $\mu $m, and that of the coupling beam is 200 $ \mu $m. There is an angle of $\theta _{s}=1.1^{\circ }$ between the coupling light and probe light. According to the time sequences shown in Fig.~\ref{fig2}a, by turning off the coupling beam, an input probe light pulse is converted to an atomic spin wave with $\mathbf{k}_{s}=\mathbf{k}_{p}-\mathbf{k}_{c}$, where $\mathbf{k}_{p}$ ($\mathbf{k}_{c}$) is the wavevector for the probe (coupling) beam. The wave number $k_{s}$ can be expressed as $k_{s}\approx k_{c}\theta _{s}$. According to the rephasing condition, $\mathbf{k}_{\pi }$ has to be in the same direction as $\mathbf{k}_{s}$, which is satisfied approximately since $\theta _{s}$ and $\theta _{\pi }$ are rather small. The time interval between the two Raman pulses has to satisfy $\Delta t/T=k_{s}/2k_{\pi }\approx \theta _{s}/2\theta _{\pi }$ where we have used $k_{c}\approx k_{1}$. Under the condition of $T=600$ $\mu $s and $\theta_{\pi }=2.1^{\circ }$, we measure the photon detection probability in the read-out mode as a function of time interval $\Delta t$. The result is shown in Fig.~\ref{fig2}c, which gives a Gaussian $1/e$ width of 46(1) $\mu $s, and an optimal interval of 154.9(5) $\mu $s. With this method the optimal intervals $\Delta t$ for different $T$ are determined, the average value of $\Delta t/T$ is calculated to be $25.8(1)\%$, which agrees very well with the theoretical estimate of $\theta _{s}/2\theta _{\pi}\approx 26.2(8)\%$. We also change the Raman angle $\theta_\pi$ for several different values, and redo the optimization process for each angle. We find that the rephasing condition is satisfied very well, as shown in Fig.~\ref{fig2}d.

\begin{figure}[hbtp]
\includegraphics[width=0.7\columnwidth]{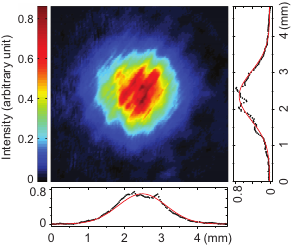}
\caption{(color online). Angular distribution of the read-out noise due to imperfection of the $\pi$ pulses. Data is measured with a CCD camera of 53 cm away from the atomic ensemble. Image center corresponds to $\theta_\pi = \theta_s$. Slight interference fringe is due to imperfection of optics used along the imaging path.}
\label{fig3}
\end{figure}

Note that when $\theta _{\pi }$ is approaching $\theta _{s}=1.1^{\circ }$,
namely $\mathbf{k}_{\pi }$ approaching $\mathbf{k}_{s}$, extremely strong
noise due to the $\pi$ pulse imperfections is observed in the probe direction.
We use a CCD camera to measure the angular distribution of this noise, with the result shown in Fig. \ref{fig3}. It suggests that the read-out noise due to imperfection of the $\pi $ pulses is highly directional, which is in conflict with both of our intuition and a previous theoretical study \cite{Heshami2011}. The angle width of this noise is calculated to be $0.28^\circ$, which corresponds to a Gaussian mode with a waist of 102 $\mu$m at the position of the atomic ensemble, which is slightly smaller than the read beam \cite{imagenote}. This highly directional noise implies that the imperfection of the $\pi$ pulses creates another collective excitation state with the wavevector $\mathbf{k}_{\pi}$. When the angle separation between $\theta_{\pi}$ and $\theta_{s}$ is much larger than the angle width in Fig. \ref{fig3}, the noise can be treated incoherently, which gives the result of $2\varepsilon N\Delta\Omega/4\pi$ in the unit of photon numbers, where $N$ is number of atoms in the mode of the coupling beam, $\varepsilon$ is the imperfection for a single $\pi$ pulse and $\Delta\Omega$ is the solid angle. When $\theta_{\pi}=\theta_{s}$, this read-out noise is collectively enhanced by a factor of $N$. Thus in order to reduce the $\pi $-pulse-induced read-out noise, the angle separation between $\theta_\pi$ and $\theta_s$ has to be large.

\begin{figure}[hbtp]
\includegraphics[width=\columnwidth]{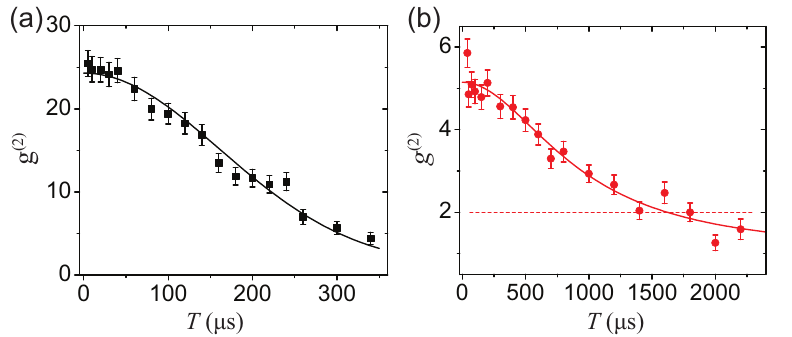}
\caption{(color online). Cross-correlation measurement as a function of storage time $T$. (a) Without applying the rephasing pulses, the lifetime is measured to be 228(6) $\protect\mu$s. (b) With the rephasing pulses applied, the lifetime is measured to be 1.20(7) ms. The reduction in $g^{(2)}$ is due to imperfection of the rephasing pulses. At $T= $ 1 ms, nonclassical correlation ($g^{(2)}>2$) is well preserved. The detection probability of the write-out photon for both measurements is set to $p_{w}=0.35\%$.}
\label{fig4}
\end{figure}

In order to directly test the feasibility of applying these rephasing pulses without destroying the single spin waves stored in the atomic ensemble, we implement the Duan-Lukin-Cirac-Zoller (DLCZ) \cite{Duan2001} protocol, for which nonclassical photon-photon correlation can be used as a criteria to verify the quantum nature of storage \cite{Kuzmich2003}. We apply a weak write pulse coupling the transition of $|g\rangle \leftrightarrow |e\rangle$ with a small detuning to induce spontaneous Raman scattering. Heralded on the detection of a single-photon in the write-out mode, a single-quanta spin-wave is created with the wavevector $\mathbf{k}_{s}=\mathbf{k}_{w}-\mathbf{k}_{wo}$, where $\mathbf{k}_{w}$ ($\mathbf{k}_{wo}$) is the wavevector of the write
beam (write-out mode). Configuration for the beam directions are shown in Fig.~\ref{fig1}. After a storage time of $T$, a strong read pulse coupling the $|s\rangle \leftrightarrow |e\rangle$ transition converts the spin wave into a single photon emitted in the read-out mode. Experimentally the cross-correlation is characterized by $g^{(2)}=p_{w,\,r}/(p_{w}\,p_{r})$ where $p_{w}$($p_{r}$) denotes the probability of detecting a write-out (read-out) photon and $p_{w,\,r}$ donates the coincidence probability between the write-out and read-out channels. $g^{(2)}>2$ means that the write-out photon and read-out photon are nonclassically correlated \cite{Felinto2005}. Without applying the rephasing pulses, the measured $g^{(2)}$ is shown in Fig.~\ref{fig4}a with $p_{\mathrm{r}}=0.28\%$ and $p_{w}=0.35\%$. The cross-correlation $g^{(2)}$ decays with a $1/e$ lifetime of 228(6) $\mu $s starting from an initial value of 24.3(6).

In order to keep away the $\pi$-pulse-induced directional noise away from the read-out mode, the intersecting angle between the Raman beams is set to be $\theta_{\pi}=1.9^{\circ}$. After optimization of the beam qualities, the fidelity of a single $\pi $ pulse is about $97\%$. The $\pi$-pulse-induced noise in the read-out mode is measured to be $p_{r}=0.8\%$. We measure the cross-correlation for a series of time points with the result shown in Fig.~\ref{fig4}b. In comparison to the case without applying the rephasing pulses, lifetime is increased by 5 folds. With the rephasing pulses on, the cross-correlation $g^{(2)}$ drops from an initial value of 5.2(1) and stays well above $2$ for about $1$ ms storage time. This result does prove that quantum nature of storage is well preserved. Higher nonclassical photon-photon correlation can be achieved by improving the accuracy of the $\pi$ pulses. We estimate that a $\pi$-pulse fidelity of $99\%$ would improve the cross-correlation to well above 10.


In summary, we have successfully managed to operate the spin echo technique in the single-quanta regime for an atomic-ensemble quantum memory. In our experiment, we find that the noise induced by slight imperfection of the $\pi$ pulses is highly directional and can be avoided by arranging the rephasing beam directions properly. With $\pi$ pulses of moderate fidelities, the quantum nature of the spin echo process is verified by observing nonclassical photon-photon correlations. We emphasize that although in our experimental demonstration we merely study the motion-induced decoherence for a cold-atomic-gas ensemble, our findings and techniques developed do apply to other decoherence mechanisms and other physical systems, like the solid-state photon-echo quantum memories \cite{Tittel2010}.

This work was supported by the National Natural Science Foundation of China, National Fundamental Research Program of China (under Grant No. 2011CB921300), and the Chinese Academy of Sciences. X.-H. B. and B. Z. acknowledge support from the Youth Qianren Program.

\textit{Note added.}$-$After completing this work we became aware of a related experiment by Jobez \emph{et al.}~\cite{Jobez2015}.

\bibliography{myref}

\begin{thebibliography}{30}
\bibitem[{ima()}]{imagenote}
\bibinfo{note}{For this measurement, beam waist for the read is set to
  $\sim$130 $\mu$m.}
\end{thebibliography}

\end{document}